\documentclass{frontiersSCNS} 
\pdfoutput=1

\def\lsim{\mathrel{\rlap{\lower4pt\hbox{\hskip1pt$\sim$}}
    \raise1pt\hbox{$<$}}}         
\def\gsim{\mathrel{\rlap{\lower4pt\hbox{\hskip1pt$\sim$}}
    \raise1pt\hbox{$>$}}}         

\usepackage[english]{babel}
\usepackage{url,lineno}

\copyrightyear{}
\pubyear{}

\def\journal{Physics}
\def\DOI{10.3389/fphy.2013.00034}
\def\articleType{arXiv:1204.2576}
\def\keyFont{\fontsize{8}{11}\helveticabold }
\def\firstAuthorLast{Boucenna {et~al.}} 
\def\Authors{S.~M. Boucenna\,$^{1}$, R.~A. Lineros\,$^{1,*}$ and J.~W.~F. Valle\,$^1$}
\def\Address{$^{1}$Instituto de F\'{\i}sica Corpuscular -- C.S.I.C./Universitat de Val{\`{e}}ncia. Val{\`e}ncia, Spain}

\def\corrAuthor{S.~M. Boucenna}
\def\corrAddress{Instituto de F\'{\i}sica Corpuscular -- C.S.I.C./Universitat de Val{\`{e}}ncia. Edificio de Institutos de Paterna, Apartado 22085, E--46071 Val{\`e}ncia, Spain}
\def\corrEmail{boucenna@ific.uv.es}

\newcommand{\figwidth}{0.8\columnwidth}
\newcommand{\sigmaA}{\langle \sigma_A v\rangle}
\newcommand{\sigmathermal}{\langle \sigma v\rangle}
\newcommand{\sigmaSI}{\sigma_{\rm SI}}

\graphicspath{{./figs/}}

\begin{document}
\onecolumn
\firstpage{1}

\title[Planck-scale effects on WIMP dark matter]{Planck-scale effects on WIMP dark matter}
\author[\firstAuthorLast ]{\Authors}
\address{}
\correspondance{}
\extraAuth{}
\extraAuth{}

\topic{Research Topic}

\maketitle

\begin{abstract}
\section{}

There exists a widely known conjecture that gravitational effects
violate global symmetries. We study the effect of global-symmetry
violating higher-dimension operators induced by Planck-scale physics
on the properties of WIMP dark matter. Using an effective description,
we show that the lifetime of the WIMP dark matter candidate can satisfy cosmological bounds
under reasonable assumptions regarding the strength of the dimension-five operators. On the other hand, the
indirect WIMP dark matter detection signal is significantly enhanced
due to new decay channels.

\tiny \keyFont{ \section{Keywords:} WIMP dark matter, decaying dark
  matter, Planck scale effects, indirect detection, particle physics }

\end{abstract}

\section{Introduction}
\label{sec:intro}

The historic discovery of neutrino oscillations and their profound
implications for neutrino properties~\citep{Maltoni:2004ei} as well as
the growing evidence for the existence of some sort of non-baryonic
dark matter~\citep{Bertone2005279} indicate the need to ammend the
Standard Model picture of matter. Here we focus on the issue of dark
matter and its stability, which is usually assumed to result from the
implementation of some \textit{ad-hoc} global discrete symmetry, like
R-parity in supersymmetric models.

It has been argued that non-perturbative gravitational effects break
all global (continuous and discrete alike)
symmetries~\citep{Giddings:1988}.
The original motivation was based on black hole physics arguments but
since then perturbative string theory has confirmed this conjecture in
various cases~\citep{Banks:2011}.
Such gravitational effects have already been considered in a number of
scenarios and models invoked to solve a number of theoretical
problems~\citep{Kamionkowski:1992mf,Holman:1992us,Akhmedov:1993,Berezinsky:1993fm,Rothstein:1993,Berezinsky:1996pb,deGouvea:2001,Masso:2004cv,Frigerio:2011}.
Here, we turn our attention to the implications of such a claim on
dark matter phenomenology, in the particular the case of weakly
interacting dark matter candidates.

Weakly Interacting Massive Particles (WIMPs) present in various
extensions of the Standard Model (SM) are among the best studied
candidates for cold DM, as they quite naturally account for the
observed relic abundance.
Indeed, their abundance at the freeze--out epoch scales as:
\begin{equation}
	\label{eq:omega}
	\Omega_{\rm CDM} h^2 \simeq 0.1 \, \frac{3 \times 10^{-26}~\rm{cm^3 s^{-1}}}{\sigmathermal_{\rm f.o.}},
\end{equation}
where $\sigmathermal_{\rm f.o.}$ is the thermal average of the
annihilation cross--section of DM times the relative velocity at
freeze--out. For weak strength interacions and DM masses in the range
from GeV to few TeV, the observed abundance is naturally obtained and
the DM candidate passes all constraints arising from Big--Bang
Nucleosynthesis (BBN), Cosmic Microwave Background (CMB) and structure
formation. This non--trivial interplay between the weak--scale physics
and the cosmological dark matter has been dubbed as the ``WIMP
miracle''.

Besides theoretical motivation and consistency, the WIMP paradigm
provides interesting direct and indirect detection prospects, through
recoil off--nuclei and annihilation to photons, neutrinos and charged
particles. As the level of sensitivity reached by direct and indirect
detection
experiments~\citep{DAMA:2008,COGENT:2011,Angloher:2011,Agnese:2013rvf,Aprile:2012nq,Akerib:2013tjd}
improves, there is growing hope that WIMP DM may be discovered
soon. Complementary searches at the LHC~\citep{Bertone:2012} and
specific indirect searches~\citep[for
example:][]{Hooper:2010mq,Weniger:2012,Fornengo:2011} may help unravel
the mystery.\\

In the absence of a protecting symmetry ensuring their stability WIMPs
are generally expected to decay. The latter is avoided in most models
by imposing in an \textit{ad--hoc} way a global symmetry (usually a
$Z_2$) that forbids the decay of the DM candidate. In this case unless
the symmetry responsible for the DM stability is local, it is expected
to be broken at short distances by gravitational effects leading to a
decaying WIMP DM. 
Dark matter will then be stable at the level of renormalizable
operators, with the Planck-mass suppressed dimension five and higher
operators responsible for its late decay. From the phenomenological
point of view, absolute stability is not a necessary condition for
being a viable DM candidate.
Instead, what is required is a lifetime larger than the current age of
the Universe $H^{-1} \approx 10^{17} {\rm s}$, where $H$ is the Hubble
constant.  Cosmic and gamma rays analysis constrain the lifetime of a
WIMP DM candidate even further to be $\tau_{\rm DM} \gsim 10^{26} {\rm
  s}\,$ ~\citep{Ibarra:2009, Ibarra:2010, Huang:2012,
  Esmaili:2012}. An important question to be addressed then is {\it
  what are the requirements on the Planck physics induced operators in
  order to preserve the validity of a WIMP DM candidate?}

In this article we will show that a sufficient requirement to ensure
an adequately long WIMP lifetime is to suppress the dimension-five
operators.  These suppressed operators will at the same time provide
us with a new source of indirect detection signals which arise from
the dark matter decay, particularly interesting in regions where
annihilation alone delivers a very faint signal. Finally, we will see
that a number of phenomenologically interesting decaying dark matter
candidates with electroweak scale masses and weak--strength couplings
fit within a WIMP paradigm, characterized by a well--controlled
cosmological history (thermal production) and viable direct detection
prospects.

We note that similar decay effects from higher scales have been analyzed before in the context of supersymmetric unified theories~\citep{Arvanitaki:2008hq} or technicolor theories~\citep{Nardi:2008ix}, where the dimension-6 GUT-suppressed operators lead to astrophysical signals that are similar to ours. However the dimension-5 operators in that case lead to too short a lifetime for DM. In contrast, the Planck scale provides a stronger suppression allowing for signals emerging from the first non-renormalizable operators.

\section{Prototype scenario}
\label{sec:model}

In order to illustrate the discussion, let us consider a very simple
prototype scheme, exhibiting generic WIMP dark matter features over a
broad phenomenological parameter range. For definiteness we assume the
Standard Model (SM) to be extended by $S_i$ ($i=1 \ldots N$) real
scalar gauge-singlets~\cite{McDonald:1994,Burgess:2000yq}. In
addition, we impose a parity symmetry to which the SM particles are
blind. Thus, the $S_i$ sector can only interact with the SM via scalar
states present in the scalar potential.  The choice of the $Z_2$
symmetry is motivated by simplicity alone, in principle any global
(discrete or continuous) symmetry that forbids the decay of the
lightest $S_i$ is a possible valid choice.

This scenario provides well known production and thermalization
mechanism via Higgs boson exchange, the so-called Higgs
portal~\citep{Davoudiasl:2005PhLB,Goudelis:2009}. The simplest model,
i.e. $N=1$, is highly constrained by current direct detection
experiments for DM mass lower than 130~GeV\footnote{There are also collider constraints for low masses (particularly from the invisible decay of the Higgs at LHC) however we do not include them here as our goal is illustrative.}. The presence of extra
gauge singlets helps overcoming these constraints thanks to
coannilihations between the LSP (lightest singlet particle) $S_1$ and
the remaining $S_{i>1}$. In this case one can reduce the value of the
cross sections involved in indirect and direct searches to an adequate
level.
In order to be efficient this process requires small enough mass
splittings~\citep{Griest:1990kh}.

In a conventional WIMP dark matter scenario our LSP would be stable
and the extra terms in the scalar potential would read as (summation
over indexes is understood):
\begin{eqnarray}
	\label{eq:potential}
	\mathcal{V}_{\rm sym} &\supset& \mu^{2}_{ij} S_i S_j + \lambda_{ij} S_i S_j (H^{\dagger}H - \frac{v^2}{2}) +  \lambda^{\prime}_{ijkl} S_i S_j S_k S_l \, ,
\end{eqnarray}
where $\mu^{2}_{ij}$, $\lambda_{ij}$ and $\lambda^{\prime}_{ijkl}$ are
the parameters of the potential. Here $H$ is the SM Higgs doublet.

\begin{figure}[tb]
	\centering
	\includegraphics[width=\figwidth]{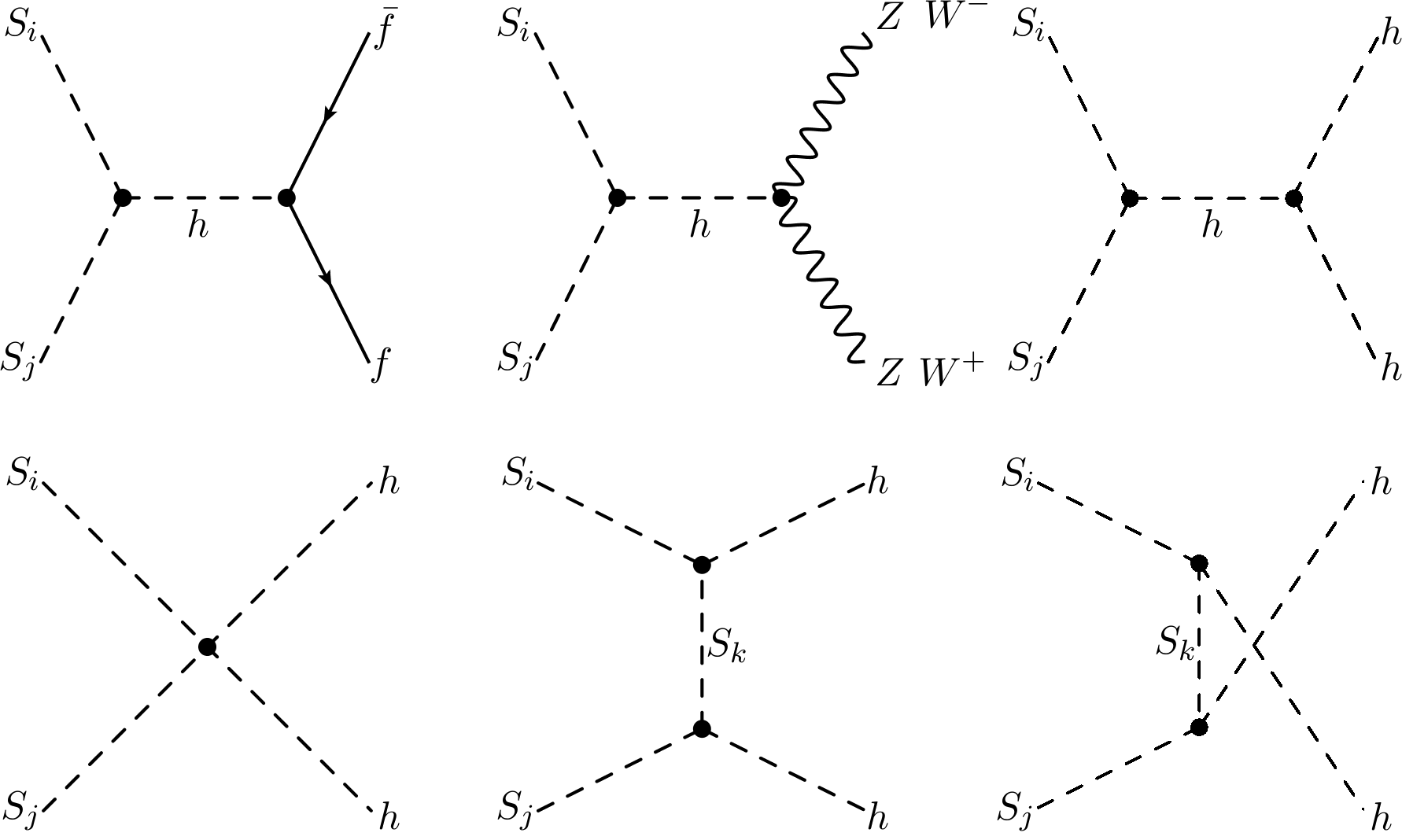}
	\caption{
	\label{fig:diag}
	Model's Feynman diagrams present in the calculation of
        the relic abundance. This model provides the relic abundance,
        annihiliation and direct detection cross sections via the
        Higgs portal.}
\end{figure}

For the sake of simplicity we consider the case $N = 2$ and suppose $\mu^{2}_{ij}$ to be diagonal with positive entries.
%
%
Equation~\ref{eq:potential} contains the relevant interaction
terms. The annihilation processes involved in the determination of the
relic abundance calculation are presented in Fig.~\ref{fig:diag}.

\section{ Parametrizing the Planck-induced effects }
\label{sec:effect-planck-induc}

As mentioned above, the global symmetry stabilizing dark matter is
likely to be broken due to the presence of gravitational effects.  In
order to implement the violation of the global symmetry responsible
for the stability of dark matter we add to $\mathcal{V}_{\rm sym}$
(Eq.~\ref{eq:potential}) an effective potential 
\begin{equation}
	\mathcal{V}_{\rm non-sym}= \sum_{n>4 , i} \frac{\kappa_n}{M_{pl}^{n-4}} \hat{\mathcal{O}}_{n,i} \,,
\end{equation}
that breaks explicitly the stabilizing symmetry through
non--renormalizable terms suppressed by powers of the Planck mass
$M_{pl}\approx 10^{19}\, {\rm GeV}$. Here $\hat{\mathcal{O}}_{n,i}$
are operators of dimension $n>4$ and $\kappa_n$ are free parameters.
In the presence of these operators induced by Planck scale effects one
has that the symmetry stabilizing DM is explicitly broken, but none of
the Standard Model symmetries. 
This leads to the decay of $S_i$, the most important of which will be the decay of the LSP.

In order to illustrate the interplay between decay and annihilation we
consider the following dimension five operators:
\begin{equation} 
	\hat{\mathcal{O}}_{5,i}^{ffh} =  Y_f  \bar{F}_{L} H f_{R} S_i \, ,
\end{equation} 
where $F_L$ is an doublet SM fermion, $f_R$ the corresponding
right-handed SU(2) singlet partner and $Y_f$ its Yukawa coupling.
This type of operators can be generated via the spontaneous breaking
of the stabilizing symmetry.  Notice that we model the gravitational
effects by parameterizing them as a scaling factor $\kappa_5$ times
the corresponding Yukawa coupling. These operators lead to a decay
lifetime $\tau_{\rm DM} = \hbar/\Gamma^{ff}$ where
\begin{equation}
	\Gamma^{ff}(M_{\rm DM}) = \sum_{f} \frac{N_c}{8 \pi^2} \left(\frac{\kappa_5 Y_f v}{M_{pl}} \right)^2 M_{\rm DM} \left(1 - \frac{4 m_f^2}{M_{\rm DM}^2} \right)^{\frac{3}{2}} \, ,
 \end{equation}
 is the dark matter decay width to fermions of mass $m_f$ and $N_c$ is
 the color number (3 for quarks and 1 for leptons). Under these
 assumptions, one has that in our prototype scenario the dark matter
 phenomenology is essentially determined by 6 parameters:
\begin{equation}
\label{eq:param}
M_{S_1}, M_{S_2}, \lambda_{11}, \lambda_{12}, \lambda_{22} \; {\rm and} \; \kappa_5.
\end{equation}

\section{WIMP dark matter annihilation and decay}
\label{sec:wimp-dark-matter}

We will now show how the mere fact of considering the global symmetry
leading to dark matter stability as an approximate one leads to extra
observational signals, in addition to the standard DM annihilation
signals. This is particularly interesting in regions of parameter
space where the latter is very faint.

The first five parameters in Eq.~(\ref{eq:param}) come from the scalar
potential (Eq.~\ref{eq:potential}).  Among these, the coupling with
the Higgs field $\lambda_{11}$ is mainly responsible for DM abundance and
direct detection signal. On the other hand $\kappa_5$
contributes to the indirect detection signal. The mass splitting
$\Delta M = M_{S_2} - M_{S_1}$ characterizes the strength of
coannihilations and their relevance for the relic abundance
calculation. When it is small enough, co-annihilations are active and
the parameter $\lambda_{12}$ controls their strength.  The coupling
$\lambda_{22}$ is the coupling of $S_2$ (the next-to-LSP) to the Higgs
scalar. It has direct impact on the dark matter relic abundance in
regions of the parameter space where both annihilations and
co-annihilations are inefficient to reproduce a good relic
abundance. In this case the latter could arise partly from the early
decays of $S_2 \to S_1$.

\begin{figure}[tb]
	\centering
	\includegraphics[width=\figwidth]{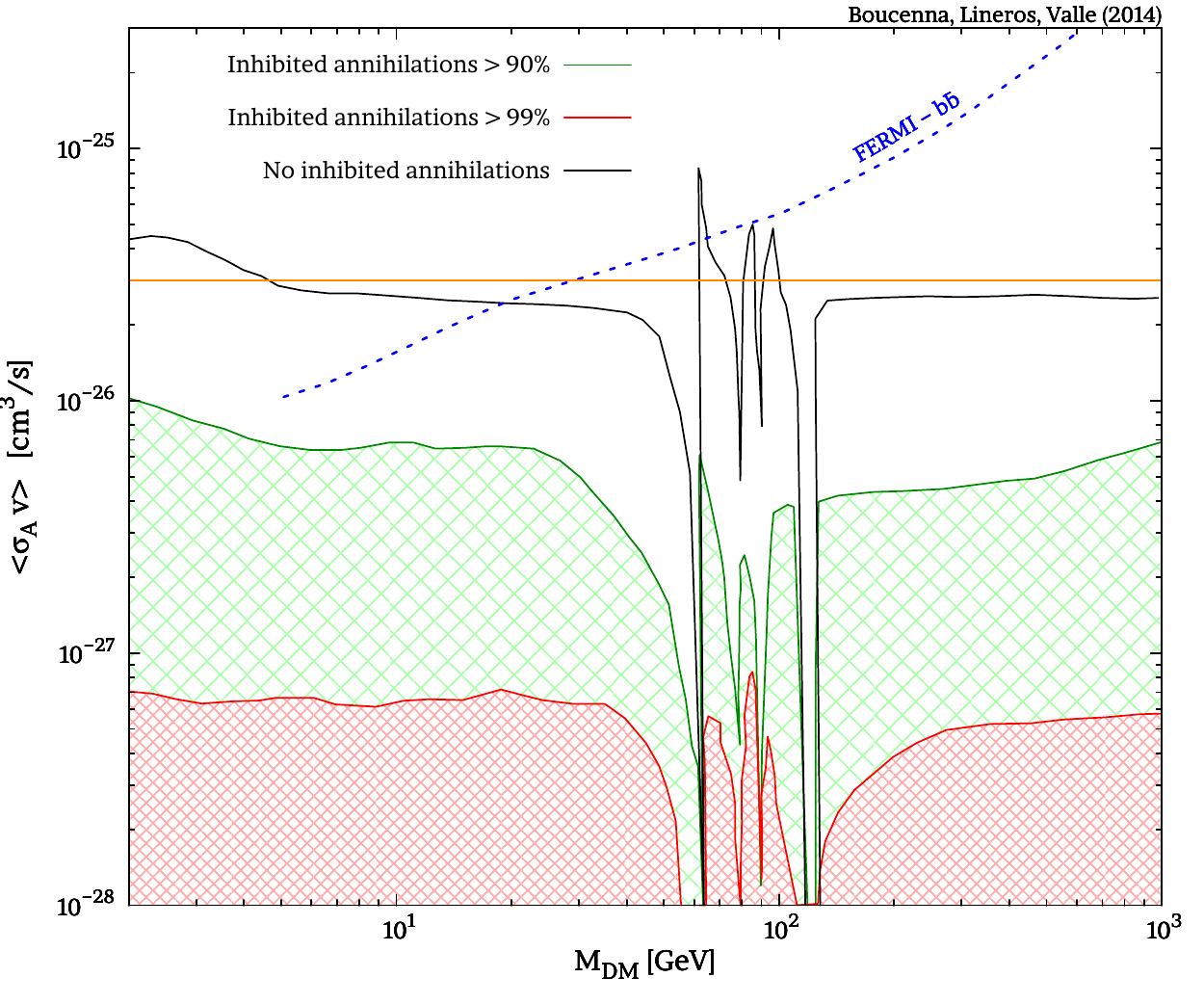}
	\caption{
	\label{fig:sv}
	Annihilation cross section $\sigmaA$ compatible with DM relic
        abundance versus DM mass.  The line shown in black corresponds
        to the simplest case of unsuppressed annihilation. The case of
        where annihilations at freeze--out are inhibited (see text)
        above 90 and 99~$\%$ is illustrated by the green and red
        shaded regions, respectively.  The thermal cross section
        $\sigmathermal_{\rm f.o.} = 3\times 10^{-26} {\rm cm}^2/{\rm s}$ and
        FERMI's constraints for annihilation into $b \bar{b}$ are also
        shown.}
\end{figure}

We present in Fig.~\ref{fig:sv} the attainable values of the thermal
average of the annihilation cross--section times velocity at present
time, $\sigmaA$, compatible with DM relic abundance
(Eq.~\ref{eq:omega}).  The couplings are varied randomly within the
limits of perturbativity, and the Higgs mass is fixed to $126 \, {\rm
  GeV}$~\citep{CMS:2012PhLB, ATLAS:2012PhLB}.
The results were obtained using the
\texttt{Micromegas}~code~\citep{Belanger:2006is,Belanger:2008sj}.
In the presence of pure annihilations, $\sigmaA$ reproduces the
expected thermal value $\sigmathermal_{\rm f.o.} \simeq
3~\times~10^{-26}~{\rm cm}^3/{\rm s}$.

Deviations from the thermal value exist in parameter regions where the
cross-section is velocity-dependent, as in the case of the
Breit--Wigner enhancement at the threshold of the Higgs
pole~\citep{Ibe:2008ye}.  

In regions where $\sigmathermal_{\rm f.o.}$ is dominated by
co-annihilation $(S_1 S_2 \to {\rm SM+SM}$ and/or $S_2 S_2 \to {\rm
  SM+SM}$) processes, the annihilation cross--section is suppressed
well below the expected thermal value.
Such ``inhibited annihilation'' regions can arise in various ways.
%
Co-annihilation is one possible mechanism, common to many dark matter models such as the minimal supersymmetric standard model two-Higgs-doublet dark matter models~\citep{Lopez:2007JC}), whose generic features are mimicked by our illustrative prototype scheme. Inhibition mechanisms have been invoked in order to obtain (or to extend) an allowed region in the parameters space~\citep{Cohen:2011ec,Edsjo:1997bg}. However, the general drawback is that the stable WIMP's indirect detection signal becomes much fainter.

\begin{figure}[tb]
	\centering
	\includegraphics[width=\figwidth]{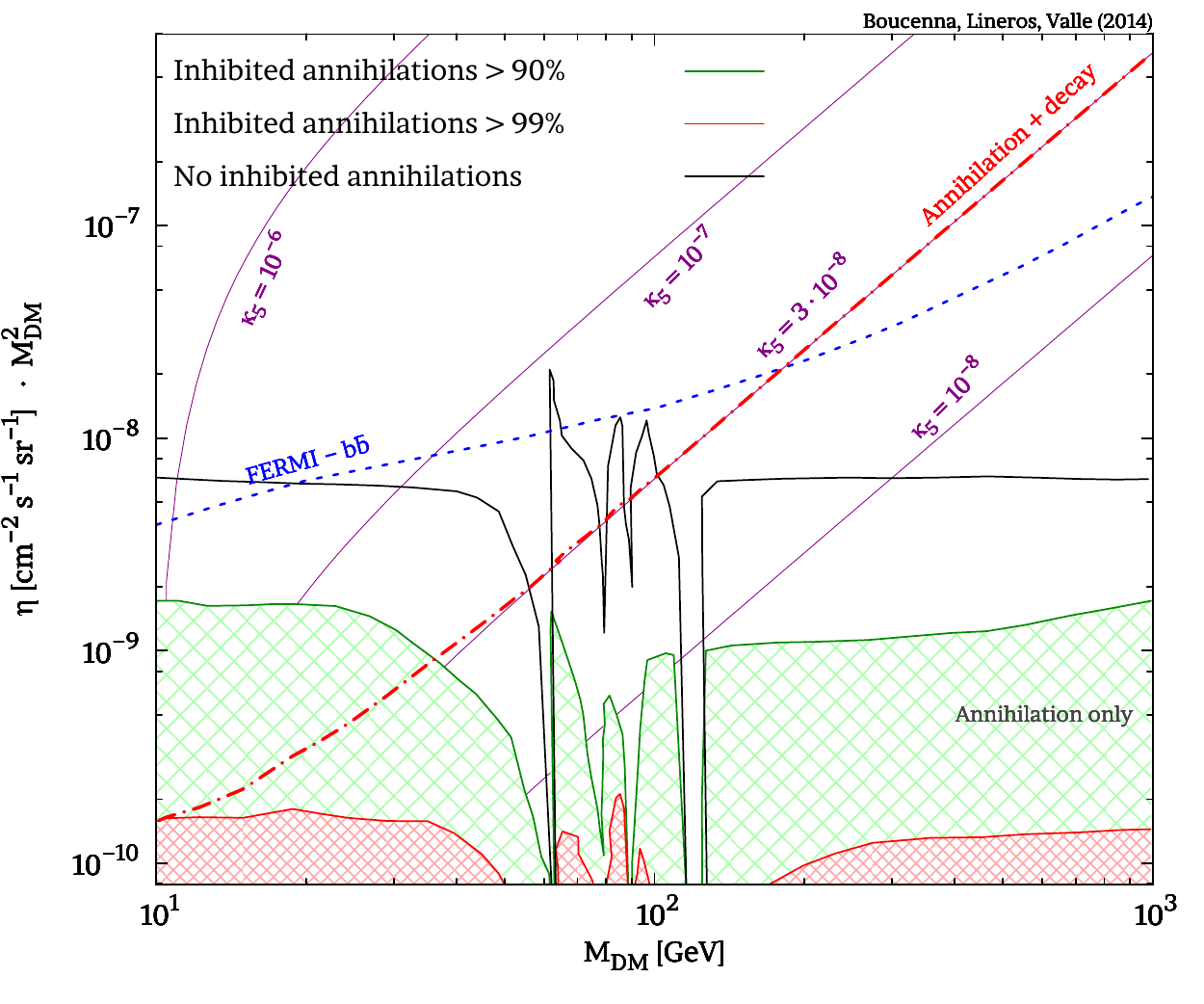}
	\caption{
	\label{fig:flux}
	The $\eta$--flux associated with dark matter decay versus DM
        mass (see text for details). We give estimates for the
        annihilation flux for the cases of 90\%, 99\%, and no
        suppression (solid green, red, and black lines respectively).
        Purple lines correspond to decay signal induced by the
        operator $\hat{\mathcal{O}}_{5}^{bbh}$ for fixed $\kappa_5$
        values. A combined signal annihilation and decay is in red
        doted--dashed line assuming $\kappa_5 = 3 \times 10^{-8}$ and
        99\% inhibited annihilation.  The bound from FERMI on
        annihilation into $b\bar{b}$~\citep{Fermi:2011} is also shown.}
\end{figure}

Now we turn to the effects of dark matter Planck--induced decays.
In Fig.~\ref{fig:flux}, we display the expected gamma--ray fluxes
arising from dark matter annihilation and decay, assuming that the
signal comes only from the production of $b \bar{b}$ through the
operator $\hat{\mathcal{O}}_{5}^{bbh}$. For illustrative purpose, we
compare with the constraints  for $\sigmaA$ from FERMI on annihilation into
$b\bar{b}$ from dwarf satellite galaxies~\citep{Fermi:2011}.
In order to compare the decay and annihilation signals we define the
$\eta$--flux:
\begin{equation}
	\eta(M_{\rm DM}) = \left\{ \begin{array}{cl} \displaystyle \frac{1}{4\pi} \frac{\sigmaA J_{\Delta \Omega}^{\rm ann}}{2 M_{\rm DM}^2} & \textnormal{ for annihilations} \\ \\
\displaystyle \frac{1}{4\pi} \frac{J_{\Delta \Omega}^{\rm dec}}{2 \tau_{\rm DM} M_{\rm DM}} & \textnormal{  for decays} \end{array} \right. \, ,
\end{equation}
where $J_{\Delta \Omega}$ is the angular averaged line of sight
integral of the DM density (squared) for decaying (annihilating) WIMP
dark matter.
In order to estimate the $\eta$--fluxes we use the $J_{\Delta \Omega}$
for galaxy clusters reported in \citet{Huang:2012}. The latter
provide weaker constraints than the analysis based on dwarf galaxies.
The observable gamma--rays flux is directly related to the
$\eta$--flux as:
\begin{equation}
	\Phi_{\gamma}(E,M_{\rm DM}) = \eta(M_{\rm DM}) \times \frac{dn_{\gamma}}{dE}(E) \, ,
\end{equation}
where ${dn_{\gamma}}/{dE}$ is the photon spectrum per single annihilation (or decay) event.
This quantity allows us to compare on the same footing both
annihilation and decay signals with FERMI constraints for an
equivalent photon spectrum. In our particular case, we compare the
signal associated to the production of $b\bar{b}$ pairs.

We notice that $\eta$--fluxes coming only from decays quickly rise with the DM mass. 
For instance, for a $50$ GeV DM mass, we would require $\kappa_5$ to
be smaller than $\sim 10^{-7}$ in order to fulfill the observational
constraints. In table~\ref{tab:bench}, we present upper bounds on
$\kappa_5$ for different DM mass values that produce DM lifetime
larger than $10^{27}$~sec, which is the current lower limit for the DM
lifetime. In contrast, inhibited annihilations (same regions as in
Fig.~\ref{fig:sv}) lead to signals that are faint and well below
observational sensitivities of indirect dark matter searches.

\begin{table}[tb]
	\centering
	\begin{tabular}{|c|c||c|}
	\hline
	\phantom{mmmmmm}$M_{\rm DM}$ [GeV] \phantom{mmmmmm}& \phantom{mmmmmm}$\kappa_5$ \phantom{mmmmmm}& most relevant dim-5 operator \\
	\hline
	5 & $\lesssim 3.5 \times 10^{-7}$ &  $\hat{\mathcal{O}}_5^{\tau \tau h}$ \\
	10 & $\lesssim 1.9 \times 10^{-7}$ & $\hat{\mathcal{O}}_5^{b b h}$ \\
	50 & $\lesssim 2.5 \times 10^{-8}$ & $\hat{\mathcal{O}}_5^{b b h}$ \\
	100 & $\lesssim 1.8 \times 10^{-8}$ & $\hat{\mathcal{O}}_5^{b b h}$ \\
	500 & $\lesssim 3.9 \times 10^{-10}$ & $\hat{\mathcal{O}}_5^{t t h}$ \\
	\hline
	\end{tabular}
\caption{\label{tab:bench} DM mass and upper bound of $\kappa_5$ for DM lifetime 
larger than $10^{27}$~sec. In addition, we present the dimension five operator which 
mainly contributes to the gamma--ray flux.}
\end{table}

Before turning to the direct detection signal, a comment on
non--thermal dark matter is in order here.

Decaying DM candidates are commonly assumed to have a non--thermal
origin, hence not to be WIMPs.  This is indeed the case of particles
with masses below or much above the weak scale. Dark matter candidates
in this class include
keV--majorons~\citep{Berezinsky:1993fm,Lattanzi:2007,Bazzocchi:2008,Esteves:2010sh,Lattanzi:2013uza},
sterile neutrinos~\citep{Dodelson:1993je}, super-heavy dark
matter~\citep{Chung:1998ua} and dark matter particles with extremely
weak couplings with Standard Model particles, such as decaying
gravitinos~\citep{Takayama:2000,Restrepo:2011rj}.

However, we see that a WIMP can be seen as purely decaying dark matter
in some regions. This implies that a number of phenomenologically
interesting decaying dark matter candidates with electroweak scale
masses and weak--strength couplings fit within a WIMP paradigm,
characterized by a well--controlled cosmological history (thermal
production) and viable direct detection prospects.

In order to ascertain the feasibility of indirect dark matter
detection one must also take into account the restrictions arising
from searches in nuclear recoil experiments.

\section{Direct detection of WIMPs}
\label{sec:direct-detection-}

\begin{figure}[tb]
	\centering
	\includegraphics[width=\figwidth]{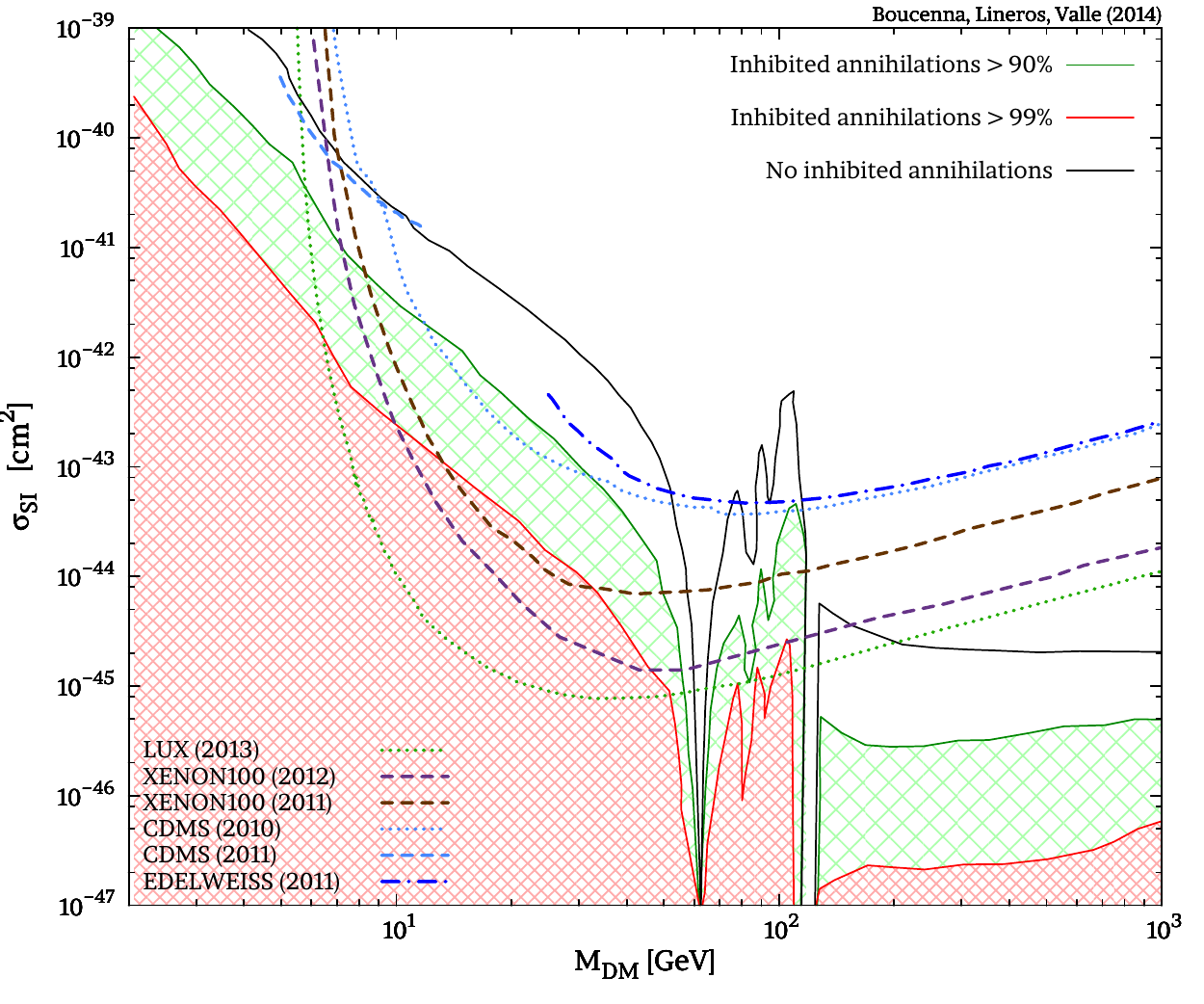}
	\caption{
	\label{fig:si}
	Spin--independent cross section versus dark matter mass.
        Exclusions from XENON100~\citep{XENON100:2011, XENON100:2012},
        CDMS~\citep{CDMS:2010, CDMS:2011}, EDELWEISS~\citep{EDELWEISS:2011} and LUX~\citep{Akerib:2013tjd} are shown. As in
        Fig.~\ref{fig:sv}, we present cases where inhibited
        annihilations produce compatible SI cross section.  WIMPs with
        masses larger than 8 GeV would produce extremely faint
        gamma--ray fluxes in order to fulfill direct detection
        constraints. }
\end{figure}

We now turn to the direct dark matter detection signal. This can be
calculated in terms of the DM mass and parametrized by $\sigmaSI$, as
shown in Fig.~\ref{fig:si}.
Note that regions with inhibited annihilations are associated with
reduced values of $\sigmaSI$. This model--dependent feature is common
to many constructions where $M_{DM}<M_W$ and annihilations are
dominated by quark final states \citep[See for a particular
case:][]{Boucenna:2011}. The correlation between $\sigmaSI$ and
$\sigmaA$ is strong and the suppression of annihilations is translated
as a weakening or absence of direct detection signal.

This correlation is certainly important to enlarge the space of
parameters allowed within the limits established by
XENON100~\citep{XENON100:2011, XENON100:2012}, CDMS~\citep{CDMS:2010,
  CDMS:2011}, EDELWEISS~\citep{EDELWEISS:2011} and
LUX~\citep{Akerib:2013tjd} experiments and opens an interesting
interplay between phenomenologically favored light WIMP DM
annihilation, decay and direct detection.

  Apart from the narrow Higgs pole region, dark matter in the
  mass range from 8 to 100 GeV, requires substantial suppression of
  the annihilation cross section, as seen in Fig.~\ref{fig:si}.
  From Fig.~\ref{fig:flux} this would preclude the indirect detection
  through the conventional annihilation signal.
  However, after including the signal originating from the decay
  modes, for ``reasonable'' choices of the $\kappa_n$ parameters
  characterizing the strength of the non-renormalizable Planck--scale
  operators, one can obtain a sizeable signal over all the mass range.
  For the case of heavy dark matter candidates
  ($>\mathcal{O}(100)\rm{GeV}$), FERMI is not sensitive enough to
  probe the thermal annihilation cross--section and decay modes
  naturally dominate the signal.
  Higher order operators are highly suppressed by the Planck scale and
  one would require unreasonably large couplings ($\sim10^{10}$) in
  order to be able to see a signal from the decay.  As a result, the
  only operators leading to a sizable decay signals are of dimension
  five.
  In other words, the stable WIMP hypothesis requires a mechanism of
  protection from Planck--effects that forbids only the dimension five
  operators.  This is in contrast to the dramatic impact upon the
  axion solution by such
  operators~\citep{Kamionkowski:1992mf,Holman:1992us}.  Alternative
  mechanisms to destabilize the DM in the context of Grand Unified
  Theories~\citep{Arvanitaki:2009Ph}, horizontal
  symmetries~\citep{Sierra:2009Ph}, instanton
  mediation~\citep{Carone:2010Ph} and kinetic mixing with hidden
  abelian gauge groups~\citep{Ibarra:2009JC} also lead to similar
  conclusions, namely, long-lived WIMPs require a high level of
  suppression of the DM decay rate.


  In summary, WIMP dark matter schemes with inhibited annihilation
  cross section have a very faint indirect detection signal. However,
  taking into account Planck-induced decay channels may re-open the
  indirect detection potential while still compatible with direct
  detection constraints.


\section{Conclusions}
\label{sec:conclusions}
In this paper, we have considered the effect of explicit gravitational
breaking of the global symmetry responsible for WIMP dark matter
stability. We find that Planck-suppressed operators of dimension six
and higher do not lead to sizable decays and can be safely neglected,
while, under reasonable assumptions, dark matter decay induced by
dimension five operators yields can be under control.

Assuming that these breaking effects are small enough to yield
sufficiently long DM decay lifetimes, there appears a genuine new
signal in regions of parameter space that were previously dark. This
illustrates the importance of taking into account the rich interplay
between direct and indirect detection of decaying WIMP dark matter.
Indeed, in models where annihilations are suppressed, the expected
signal of indirect detection is very faint which also leads to a small
direct detection signal in most cases. However, including the WIMP
decays induced by the Planck--scale effects, one has an extra source
of signal which makes the WIMP dark matter detectable.

On the other hand, a large class of decaying dark matter candidates,
with masses in the WIMP range and weak--strength couplings, naturally
accommodates a WIMP thermal production scenario.

Finally, an unambiguous detection of a mixed decay and annihilation
signal may offer a very interesting window into Planck--scale physics,
opening an unexpected phenomenological implication of the WIMP
paradigm.

\section*{Acknowledgments} 

We thank N. Fornengo and M. Taoso for their comments and suggestions
upon reading the manuscript. This work was supported by the Spanish
MINECO under grants FPA2011-22975 and MULTIDARK CSD2009-00064
(Consolider-Ingenio 2010 Programme), by Prometeo/2009/091 (Generalitat
Valenciana), by the EU ITN UNILHC PITN-GA-2009-237920.

\newcommand{\prd}{PRD}
\newcommand{\jcap}{JCAP}

\bibliographystyle{frontiersinSCNS&ENG}
\bibliography{BLV_refs,merged,newrefs}

\begin{thebibliography}{61}
\providecommand{\natexlab}[1]{#1}
\expandafter\ifx\csname urlstyle\endcsname\relax
  \providecommand{\doi}[1]{doi:\discretionary{}{}{}#1}\else
  \providecommand{\doi}{doi:\discretionary{}{}{}\begingroup
  \urlstyle{rm}\Url}\fi
\providecommand{\selectlanguage}[1]{\relax}

\bibitem[{\textbf{{Aalseth} et~al.}(2011)}]{COGENT:2011}
{Aalseth}, C.~E. et~al. (2011), {Results from a Search for Light-Mass Dark
  Matter with a p-Type Point Contact Germanium Detector}, \emph{Physical Review
  Letters}, 106, 13, 131301, \doi{10.1103/PhysRevLett.106.131301}

\bibitem[{\textbf{{Ackermann} et~al.}(2011)}]{Fermi:2011}
{Ackermann}, M. et~al. (2011), {Constraining Dark Matter Models from a Combined
  Analysis of Milky Way Satellites with the Fermi Large Area Telescope},
  \emph{PRL}, 107, 24, 241302, \doi{10.1103/PhysRevLett.107.241302}

\bibitem[{\textbf{Agnese et~al.}(2013)}]{Agnese:2013rvf}
Agnese, R. et~al. (2013), {Dark Matter Search Results Using the Silicon
  Detectors of CDMS II}, \emph{Phys.Rev.Lett.}

\bibitem[{\textbf{{Ahmed} et~al.}(2011)}]{CDMS:2011}
{Ahmed}, Z. et~al. (2011), {Results from a Low-Energy Analysis of the CDMS II
  Germanium Data}, \emph{Physical Review Letters}, 106, 13, 131302,
  \doi{10.1103/PhysRevLett.106.131302}

\bibitem[{\textbf{Akerib et~al.}(2013)}]{Akerib:2013tjd}
Akerib, D. et~al. (2013), {First results from the LUX dark matter experiment at
  the Sanford Underground Research Facility}

\bibitem[{\textbf{{Akhmedov} et~al.}(1993)}]{Akhmedov:1993}
{Akhmedov}, E.~K. et~al. (1993), {Planck scale effects on the majoron},
  \emph{Physics Letters B}, 299, 90--93, \doi{10.1016/0370-2693(93)90887-N}

\bibitem[{\textbf{{Angloher} et~al.}(2012)}]{Angloher:2011}
{Angloher}, G. et~al. (2012), Results from 730 kg days of the cresst-ii dark
  matter search, \emph{European Physical Journal C}, 72, 1971,
  \doi{10.1140/epjc/s10052-012-1971-8}

\bibitem[{\textbf{{Aprile} et~al.}(2011)}]{XENON100:2011}
{Aprile}, E. et~al. (2011), {Dark Matter Results from 100 Live Days of XENON100
  Data}, \emph{Physical Review Letters}, 107, 13, 131302,
  \doi{10.1103/PhysRevLett.107.131302}

\bibitem[{\textbf{Aprile et~al.}(2012)}]{Aprile:2012nq}
Aprile, E. et~al. (2012), {Dark Matter Results from 225 Live Days of XENON100
  Data}, \emph{Phys.Rev.Lett.}, 109, 181301,
  \doi{10.1103/PhysRevLett.109.181301}

\bibitem[{\textbf{Arvanitaki et~al.}(2009)\textbf{Arvanitaki, Dimopoulos,
  Dubovsky, Graham, Harnik et~al.}}]{Arvanitaki:2008hq}
Arvanitaki, A., Dimopoulos, S., Dubovsky, S., Graham, P.~W., Harnik, R., et~al.
  (2009), {Astrophysical Probes of Unification}, \emph{Phys.Rev.}, D79, 105022,
  \doi{10.1103/PhysRevD.79.105022}

\bibitem[{\textbf{{Arvanitaki} et~al.}(2009)}]{Arvanitaki:2009Ph}
{Arvanitaki}, A. et~al. (2009), {Astrophysical probes of unification},
  \emph{\prd}, 79, 10, 105022, \doi{10.1103/PhysRevD.79.105022}

\bibitem[{\textbf{{ATLAS Collaboration}}(2012)}]{ATLAS:2012PhLB}
{ATLAS Collaboration} (2012), {Observation of a new particle in the search for
  the Standard Model Higgs boson with the ATLAS detector at the LHC},
  \emph{Physics Letters B}, 716, 1--29, \doi{10.1016/j.physletb.2012.08.020}

\bibitem[{\textbf{{Banks} and {Seiberg}}(2011)}]{Banks:2011}
{Banks}, T. and {Seiberg}, N. (2011), {Symmetries and strings in field theory
  and gravity}, \emph{PRD}, 83, 8, 084019, \doi{10.1103/PhysRevD.83.084019}

\bibitem[{\textbf{{Bazzocchi} et~al.}(2008)}]{Bazzocchi:2008}
{Bazzocchi}, F. et~al. (2008), {X-ray photons from late-decaying majoron dark
  matter}, \emph{JCAP}, 8, 13, \doi{10.1088/1475-7516/2008/08/013}

\bibitem[{\textbf{Belanger et~al.}(2007)}]{Belanger:2006is}
Belanger, G. et~al. (2007), {MicrOMEGAs 2.0: A Program to calculate the relic
  density of dark matter in a generic model}, \emph{Comput.Phys.Commun.}, 176,
  367--382, \doi{10.1016/j.cpc.2006.11.008}

\bibitem[{\textbf{Belanger et~al.}(2009)}]{Belanger:2008sj}
Belanger, G. et~al. (2009), {Dark matter direct detection rate in a generic
  model with micrOMEGAs 2.2}, \emph{Comput.Phys.Commun.}, 180, 747--767,
  \doi{10.1016/j.cpc.2008.11.019}

\bibitem[{\textbf{Berezinsky et~al.}(1998)\textbf{Berezinsky, Joshipura, and
  Valle}}]{Berezinsky:1996pb}
Berezinsky, V., Joshipura, A.~S., and Valle, J. (1998), {Gravitational
  violation of R-parity and its cosmological signatures}, \emph{Phys.Rev.},
  D57, 147--151, \doi{10.1103/PhysRevD.57.147}

\bibitem[{\textbf{Berezinsky and Valle}(1993)}]{Berezinsky:1993fm}
Berezinsky, V. and Valle, J. (1993), {The KeV majoron as a dark matter
  particle}, \emph{Phys.Lett.}, B318, 360--366,
  \doi{10.1016/0370-2693(93)90140-D}

\bibitem[{\textbf{{Bernabei} et~al.}(2008)}]{DAMA:2008}
{Bernabei}, R. et~al. (2008), {First results from DAMA/LIBRA and the combined
  results with DAMA/NaI}, \emph{European Physical Journal C}, 56, 333,
  \doi{10.1140/epjc/s10052-008-0662-y}

\bibitem[{\textbf{Bertone et~al.}(2005)\textbf{Bertone, Hooper, and
  Silk}}]{Bertone2005279}
Bertone, G., Hooper, D., and Silk, J. (2005), Particle dark matter: evidence,
  candidates and constraints, \emph{Physics Reports}, 405, 5–6, 279 -- 390,
  \doi{10.1016/j.physrep.2004.08.031}

\bibitem[{\textbf{{Bertone} et~al.}(2012)}]{Bertone:2012}
{Bertone}, G. et~al. (2012), {Complementarity of indirect and accelerator dark
  matter searches}, \emph{\prd}, 85, 5, 055014,
  \doi{10.1103/PhysRevD.85.055014}

\bibitem[{\textbf{{Boucenna} and {Profumo}}(2011)}]{Boucenna:2011}
{Boucenna}, M.~S. and {Profumo}, S. (2011), {Direct and indirect singlet scalar
  dark matter detection in the lepton-specific two-Higgs-doublet model},
  \emph{PRD}, 84, 5, 055011, \doi{10.1103/PhysRevD.84.055011}

\bibitem[{\textbf{Burgess et~al.}(2001)\textbf{Burgess, Pospelov, and ter
  Veldhuis}}]{Burgess:2000yq}
Burgess, C.~P., Pospelov, M., and ter Veldhuis, T. (2001), {The minimal model
  of nonbaryonic dark matter: A singlet scalar}, \emph{Nucl. Phys.}, B619,
  709--728, \doi{10.1016/S0550-3213(01)00513-2}

\bibitem[{\textbf{{Carone} et~al.}(2010)\textbf{{Carone}, {Erlich}, and
  {Primulando}}}]{Carone:2010Ph}
{Carone}, C.~D., {Erlich}, J., and {Primulando}, R. (2010), {Decaying dark
  matter from dark instantons}, \emph{\prd}, 82, 5, 055028,
  \doi{10.1103/PhysRevD.82.055028}

\bibitem[{\textbf{{CDMS II Collaboration}}(2010)}]{CDMS:2010}
{CDMS II Collaboration} (2010), {Dark Matter Search Results from the CDMS II
  Experiment}, \emph{Science}, 327, 1619--, \doi{10.1126/science.1186112}

\bibitem[{\textbf{Chung et~al.}(1998)\textbf{Chung, Kolb, and
  Riotto}}]{Chung:1998ua}
Chung, D.~J., Kolb, E.~W., and Riotto, A. (1998), {Nonthermal supermassive dark
  matter}, \emph{Phys.Rev.Lett.}, 81, 4048--4051,
  \doi{10.1103/PhysRevLett.81.4048}

\bibitem[{\textbf{{CMS Collaboration}}(2012)}]{CMS:2012PhLB}
{CMS Collaboration} (2012), {Observation of a new boson at a mass of 125 GeV
  with the CMS experiment at the LHC}, \emph{Physics Letters B}, 716, 30--61,
  \doi{10.1016/j.physletb.2012.08.021}

\bibitem[{\textbf{{Cohen} et~al.}(2012)}]{Cohen:2011ec}
{Cohen}, T. et~al. (2012), {Singlet-doublet dark matter}, \emph{\prd}, 85, 7,
  075003, \doi{10.1103/PhysRevD.85.075003}

\bibitem[{\textbf{{Davoudiasl} et~al.}(2005)}]{Davoudiasl:2005PhLB}
{Davoudiasl}, H. et~al. (2005), {The new Minimal Standard Model}, \emph{Physics
  Letters B}, 609, 117--123, \doi{10.1016/j.physletb.2005.01.026}

\bibitem[{\textbf{{de Gouv{\^e}a} and {Valle}}(2001)}]{deGouvea:2001}
{de Gouv{\^e}a}, A. and {Valle}, J.~W.~F. (2001), {Minimalistic neutrino mass
  model}, \emph{Physics Letters B}, 501, 115--127,
  \doi{10.1016/S0370-2693(01)00103-4}

\bibitem[{\textbf{Dodelson and Widrow}(1994)}]{Dodelson:1993je}
Dodelson, S. and Widrow, L.~M. (1994), {Sterile-neutrinos as dark matter},
  \emph{Phys.Rev.Lett.}, 72, 17--20, \doi{10.1103/PhysRevLett.72.17}

\bibitem[{\textbf{{EDELWEISS Collaboration}}(2011)}]{EDELWEISS:2011}
{EDELWEISS Collaboration} (2011), {Final results of the EDELWEISS-II WIMP
  search using a 4-kg array of cryogenic germanium detectors with interleaved
  electrodes}, \emph{Physics Letters B}, 702, 329--335,
  \doi{10.1016/j.physletb.2011.07.034}

\bibitem[{\textbf{Edsjo and Gondolo}(1997)}]{Edsjo:1997bg}
Edsjo, J. and Gondolo, P. (1997), {Neutralino relic density including
  coannihilations}, \emph{Phys.Rev.}, D56, 1879--1894,
  \doi{10.1103/PhysRevD.56.1879}

\bibitem[{\textbf{{Esmaili} et~al.}(2012)\textbf{{Esmaili}, {Ibarra}, and
  {Peres}}}]{Esmaili:2012}
{Esmaili}, A., {Ibarra}, A., and {Peres}, O.~L.~G. (2012), {Probing the
  stability of superheavy dark matter particles with high-energy neutrinos},
  \emph{\jcap}, 11, 034, \doi{10.1088/1475-7516/2012/11/034}

\bibitem[{\textbf{Esteves et~al.}(2010)}]{Esteves:2010sh}
Esteves, J. et~al. (2010), {$A_4$-based neutrino masses with Majoron decaying
  dark matter}, \emph{Phys.Rev.}, D82, 073008, \doi{10.1103/PhysRevD.82.073008}

\bibitem[{\textbf{{Fornengo} et~al.}(2011)\textbf{{Fornengo}, {Lineros},
  {Regis}, and {Taoso}}}]{Fornengo:2011}
{Fornengo}, N., {Lineros}, R., {Regis}, M., and {Taoso}, M. (2011),
  {Possibility of a Dark Matter Interpretation for the Excess in Isotropic
  Radio Emission Reported by ARCADE}, \emph{Physical Review Letters}, 107, 26,
  271302, \doi{10.1103/PhysRevLett.107.271302}

\bibitem[{\textbf{{Frigerio} et~al.}(2011)\textbf{{Frigerio}, {Hambye}, and
  {Masso}}}]{Frigerio:2011}
{Frigerio}, M., {Hambye}, T., and {Masso}, E. (2011), {Sub-GeV Dark Matter as
  Pseudo-Nambu-Goldstone Bosons from the Seesaw Scale}, \emph{Physical Review
  X}, 1, 2, 021026, \doi{10.1103/PhysRevX.1.021026}

\bibitem[{\textbf{{Giddings} and {Strominger}}(1988)}]{Giddings:1988}
{Giddings}, S.~B. and {Strominger}, A. (1988), {Loss of incoherence and
  determination of coupling constants in quantum gravity}, \emph{Nuclear
  Physics B}, 307, 854--866, \doi{10.1016/0550-3213(88)90109-5}

\bibitem[{\textbf{{Goudelis} et~al.}(2009)\textbf{{Goudelis}, {Mambrini}, and
  {Yaguna}}}]{Goudelis:2009}
{Goudelis}, A., {Mambrini}, Y., and {Yaguna}, C. (2009), {Antimatter signals of
  singlet scalar dark matter}, \emph{\jcap}, 12, 008,
  \doi{10.1088/1475-7516/2009/12/008}

\bibitem[{\textbf{Griest and Seckel}(1991)}]{Griest:1990kh}
Griest, K. and Seckel, D. (1991), {Three exceptions in the calculation of relic
  abundances}, \emph{Phys.Rev.}, D43, 3191--3203,
  \doi{10.1103/PhysRevD.43.3191}

\bibitem[{\textbf{Holman et~al.}(1992)}]{Holman:1992us}
Holman, R. et~al. (1992), {Solutions to the strong CP problem in a world with
  gravity}, \emph{Phys.Lett.}, B282, 132--136,
  \doi{10.1016/0370-2693(92)90491-L}

\bibitem[{\textbf{{Hooper} and {Goodenough}}(2011)}]{Hooper:2010mq}
{Hooper}, D. and {Goodenough}, L. (2011), {Dark matter annihilation in the
  Galactic Center as seen by the Fermi Gamma Ray Space Telescope},
  \emph{Physics Letters B}, 697, 412--428, \doi{10.1016/j.physletb.2011.02.029}

\bibitem[{\textbf{{Huang} et~al.}(2012)\textbf{{Huang}, {Vertongen}, and
  {Weniger}}}]{Huang:2012}
{Huang}, X., {Vertongen}, G., and {Weniger}, C. (2012), {Probing dark matter
  decay and annihilation with Fermi LAT observations of nearby galaxy
  clusters}, \emph{JCAP}, 1, 42, \doi{10.1088/1475-7516/2012/01/042}

\bibitem[{\textbf{{Ibarra} and {Tran}}(2009)}]{Ibarra:2009}
{Ibarra}, A. and {Tran}, D. (2009), {Decaying dark matter and the PAMELA
  anomaly}, \emph{JCAP}, 2, 21, \doi{10.1088/1475-7516/2009/02/021}

\bibitem[{\textbf{{Ibarra} et~al.}(2010)\textbf{{Ibarra}, {Tran}, and
  {Weniger}}}]{Ibarra:2010}
{Ibarra}, A., {Tran}, D., and {Weniger}, C. (2010), {Decaying dark matter in
  light of the PAMELA and Fermi LAT data}, \emph{JCAP}, 1, 9,
  \doi{10.1088/1475-7516/2010/01/009}

\bibitem[{\textbf{{Ibarra} et~al.}(2009)}]{Ibarra:2009JC}
{Ibarra}, A. et~al. (2009), {Cosmic rays from leptophilic dark matter decay via
  kinetic mixing}, \emph{\jcap}, 8, 017, \doi{10.1088/1475-7516/2009/08/017}

\bibitem[{\textbf{Ibe et~al.}(2009)\textbf{Ibe, Murayama, and
  Yanagida}}]{Ibe:2008ye}
Ibe, M., Murayama, H., and Yanagida, T. (2009), {Breit-Wigner Enhancement of
  Dark Matter Annihilation}, \emph{Phys.Rev.}, D79, 095009,
  \doi{10.1103/PhysRevD.79.095009}

\bibitem[{\textbf{Kamionkowski and March-Russell}(1992)}]{Kamionkowski:1992mf}
Kamionkowski, M. and March-Russell, J. (1992), {Planck scale physics and the
  Peccei-Quinn mechanism}, \emph{Phys.Lett.}, B282, 137--141,
  \doi{10.1016/0370-2693(92)90492-M}

\bibitem[{\textbf{Lattanzi et~al.}(2013)\textbf{Lattanzi, Riemer-Sorensen,
  Tortola, and Valle}}]{Lattanzi:2013uza}
Lattanzi, M., Riemer-Sorensen, S., Tortola, M., and Valle, J. W.~F. (2013),
  {Updated CMB, X- and gamma-ray constraints on majoron dark matter},
  \emph{Phys.Rev.}, D88, 063528, \doi{10.1103/PhysRevD.88.063528}

\bibitem[{\textbf{{Lattanzi} and {Valle}}(2007)}]{Lattanzi:2007}
{Lattanzi}, M. and {Valle}, J.~W.~F. (2007), {Decaying Warm Dark Matter and
  Neutrino Masses}, \emph{Physical Review Letters}, 99, 12, 121301,
  \doi{10.1103/PhysRevLett.99.121301}

\bibitem[{\textbf{{Lopez Honorez} et~al.}(2007)}]{Lopez:2007JC}
{Lopez Honorez}, L. et~al. (2007), {The inert doublet model: an archetype for
  dark matter}, \emph{\jcap}, 2, 028, \doi{10.1088/1475-7516/2007/02/028}

\bibitem[{\textbf{Maltoni et~al.}(2004)\textbf{Maltoni, Schwetz, Tortola, and
  Valle}}]{Maltoni:2004ei}
Maltoni, M., Schwetz, T., Tortola, M., and Valle, J. (2004), {Status of global
  fits to neutrino oscillations}, \emph{New J.Phys.}, 6, 122,
  \doi{10.1088/1367-2630/6/1/122}, this review gives a comprehensive set of
  references

\bibitem[{\textbf{Masso et~al.}(2004)\textbf{Masso, Rota, and
  Zsembinszki}}]{Masso:2004cv}
Masso, E., Rota, F., and Zsembinszki, G. (2004), {Planck-scale effects on
  global symmetries: Cosmology of pseudo-Goldstone bosons}, \emph{Phys.Rev.},
  D70, 115009, \doi{10.1103/PhysRevD.70.115009}

\bibitem[{\textbf{{McDonald}}(1994)}]{McDonald:1994}
{McDonald}, J. (1994), {Gauge singlet scalars as cold dark matter}, \emph{PRD},
  50, 3637--3649, \doi{10.1103/PhysRevD.50.3637}

\bibitem[{\textbf{Nardi et~al.}(2009)\textbf{Nardi, Sannino, and
  Strumia}}]{Nardi:2008ix}
Nardi, E., Sannino, F., and Strumia, A. (2009), {Decaying Dark Matter can
  explain the e+- excesses}, \emph{JCAP}, 0901, 043,
  \doi{10.1088/1475-7516/2009/01/043}

\bibitem[{\textbf{{Restrepo} et~al.}(2012)}]{Restrepo:2011rj}
{Restrepo}, D. et~al. (2012), {Gravitino dark matter and neutrino masses with
  bilinear R-parity violation}, \emph{\prd}, 85, 2, 023523,
  \doi{10.1103/PhysRevD.85.023523}

\bibitem[{\textbf{{Rothstein} et~al.}(1993)\textbf{{Rothstein}, {Babu}, and
  {Seckel}}}]{Rothstein:1993}
{Rothstein}, I.~Z., {Babu}, K.~S., and {Seckel}, D. (1993), {Planck scale
  symmetry breaking and majoron physics}, \emph{Nuclear Physics B}, 403,
  725--748, \doi{10.1016/0550-3213(93)90368-Y}

\bibitem[{\textbf{{Sierra} et~al.}(2009)\textbf{{Sierra}, {Restrepo}, and
  {Zapata}}}]{Sierra:2009Ph}
{Sierra}, D.~A., {Restrepo}, D., and {Zapata}, O. (2009), {Decaying neutralino
  dark matter in anomalous U(1)$_{H}$ models}, \emph{\prd}, 80, 5, 055010,
  \doi{10.1103/PhysRevD.80.055010}

\bibitem[{\textbf{{Takayama} and {Yamaguchi}}(2000)}]{Takayama:2000}
{Takayama}, F. and {Yamaguchi}, M. (2000), {Gravitino dark matter without
  R-parity}, \emph{Physics Letters B}, 485, 388--392,
  \doi{10.1016/S0370-2693(00)00726-7}

\bibitem[{\textbf{{Weniger}}(2012)}]{Weniger:2012}
{Weniger}, C. (2012), {A tentative gamma-ray line from Dark Matter annihilation
  at the Fermi Large Area Telescope}, \emph{\jcap}, 8, 007,
  \doi{10.1088/1475-7516/2012/08/007}

\bibitem[{\textbf{{XENON100 Collaboration}}(2012)}]{XENON100:2012}
{XENON100 Collaboration} (2012), {Dark Matter Results from 225 Live Days of
  XENON100 Data}, \emph{Physical Review Letters}, 109, 18, 181301,
  \doi{10.1103/PhysRevLett.109.181301}

\end{thebibliography}

\end{document}